\documentclass[preprint,aps,12pt,showpacs,nofootinbib,tightenlines,amsmath,amssymb]{revtex4}
\usepackage{amsmath}
\usepackage{graphicx}
\usepackage{amssymb}
\usepackage{color}
\usepackage{slashed}
\newcommand{\VEV}[1]{\left\langle #1\right\rangle}

\newcommand{\p}{\partial}

\newcommand{\MeV}{\;\text{MeV}}

\begin{document}
\def\intdk{\int\frac{d^4k}{(2\pi)^4}}
\def\sla{\hspace{-0.17cm}\slash}
\hfill

\title{IR-improved Soft-wall AdS/QCD Model for Baryons }

\author{Zhen Fang}\email{fangzhen@itp.ac.cn}

\author{Dan-Ning Li}\email{lidn@itp.ac.cn}

\author{Yue-Liang Wu}\email{ylwu@itp.ac.cn}

\affiliation{State Key Laboratory of Theoretical Physics(SKLTP)\\
Kavli Institute for Theoretical Physics China (KITPC) \\
Institute of Theoretical Physics, Chinese Academy of Sciences, Beijing 100190 \\
University of Chinese Academy of Sciences (UCAS), P. R. China }

\date{\today}

\begin{abstract}
We construct an infrared-improved soft-wall AdS/QCD model for baryons by considering the infrared-modified 5D conformal mass and Yukawa coupling of the bulk baryon field. The model is also built by taking into account the parity-doublet pattern for the excited baryons. When taking the bulk vacuum structure of the meson field to be the one obtained consistently in the infrared-improved soft-wall AdS/QCD model for mesons, we arrive at a consistent prediction for the baryon mass spectrum in even and odd parity. The prediction shows a remarkable agreement with the experimental data. We also perform a calculation for the $\rho(a_1)$ meson-nucleon coupling constant and obtain a consistent result in comparison with the experimental data and many other models.
\end{abstract}

\maketitle

\section{Introduction}
\label{Chap:Intro}

The quantum chromodynamics (QCD) is a fundamental theory of strong force for describing the strong interactions among quarks and gluons.  QCD has successfully been applied to study the perturbative effects of strong interactions at high energies due to the property of asymptotic freedom\cite{Gross:1973id,Politzer:1973fx}. However, it is not solvable analytically in the nonperturbative low-energy regime. Many effective theories or tools have been constructed to give a low-energy description of QCD, such as the chiral perturbation theory or lattice QCD. The discovery of AdS/CFT \cite{Maldacena:1997re,Gubser:1998bc,Witten:1998qj}, which establishes the duality between the weak coupled supergravity in $AdS_5$ and the strong coupled $N = 4$ super Yang-Mills gauge theory in the boundary, supply new idea for solving the challenge problems of strong interaction at low energies. Based on AdS/CFT, a holographic description of QCD has been established. A number of researches have been focused on the low-energy phenomenology of hadrons, they can be divided into top-down and bottom-up approach according to the bias of which side in the duality between the gravity part and the gauge field part. The former starts from some brane configurations in string theory to reproduce some basic features of QCD \cite{Kruczenski:2003uq,Sakai:2004cn}. The latter is constructed on the basis of low-energy properties of QCD, such as chiral symmetry breaking and quark confinement \cite{Erlich:2005qh,Karch:2006pv,BT1}.

There are hard-wall and soft-wall models in the bottom-up approach. The hard-wall model \cite{Erlich:2005qh} used a sharp cut-off in the extra dimension to realize the confinement and chiral symmetry breaking, while the mass spectrum of excited mesons in this model cannot match the experimental data well. In the motivated soft-wall model \cite{Karch:2006pv}, an IR-suppressed dilaton term was added to reproduce the Regge behavior of mass spectrum, but the chiral symmetry breaking phenomenon cannot be consistently realized. Many efforts \cite{soft wall 1,Gherghetta:2009ac,Sui:2009xe,Li:2012ay,Cui:2013xva} have been made to build more realistic and predictive models, so that the mass spectrum of mesons can be matched up better in comparison with the experimental data, some other characteristic quantities, such as interacting couplings or form factors, can also be produced well. In \cite{Cui:2013xva}, we proposed a modified soft-wall AdS/QCD model which can lead to a more consistent prediction for the mass spectrum of mesons.


Applications of AdS/QCD approach to baryons have attracted much attention \cite{deTeramond:2005su,Hong:2006ta,Kim:2007xi,Forkel:2007cm,Vega:2008te,Pomarol:2008aa,Abidin:2009hr,Vega:2010ns,Zhang:2010tk}. A number of quantities including the baryon spectrum and electromagnetic form factors of nulceons as well as the meson-nucleon couplings have been calculated in either hard-wall or soft-wall models. In \cite{Hong:2006ta}, a hard-wall model for low-lying spin-$\frac{1}{2}$ baryons that includes the effects of chiral symmetry breaking was proposed to explain the parity-doublet pattern of excited baryons.
Some other efforts have been made to tackle the problems in the soft-wall AdS/QCD model for baryons\cite{Gutsche:2011vb,Gutsche:2012bp}, while it can be shown that there exists a massless mode due to the lack of chiral symmetry breaking. In all these models, the baryon spectrum cannot be well matched up with the experimentally measured values.

In this paper, we shall pay attention to the infrared(IR) behavior of AdS/QCD model and build an IR-improved soft-wall AdS/QCD model for baryons. The model is constructed by considering the IR-modified conformal mass and Yukawa coupling of the bulk baryon field with the bulk meson field. To be consistent with the soft-wall AdS/QCD model for mesons, the bulk vacuum structure for the bulk meson field is taken to be the same as the one obtained in the IR-improved soft-wall AdS/QCD model for mesons which has been shown to provide a consistent prediction for the mass spectra of resonance mesons\cite{Cui:2013xva}. The model is built by taking into account the parity-doublet pattern of excited baryons as discussed in the hard-wall baryon model\cite{Hong:2006ta}. As a consequence, we arrive at a more realistic IR-improved soft-wall AdS/QCD model for baryons. Such a model can provide a consistent prediction for the baryon mass spectrum which fits to the experimental data very well. We also perform a calculation for the $\rho(a_1)$ meson-nucleon coupling constant and compare it with the results obtained from the empirical estimation and other models. The conclusions and remarks are presented in the last part.

\section{baryon mass spectrum in IR-improved soft-wall AdS/QCD model}

\subsection{Action of IR-improved Soft-wall AdS/QCD Model}

Let us first construct the soft-wall AdS/QCD model for baryons. We use the standard 5D AdS space-time as the background
\begin{equation}\label{metric}
ds^2=e^{2A(z)}\left(\eta_{\mu\nu}dx^{\mu}dx^{\nu}-dz^2\right);\qquad A(z)=-\mathrm{ln}\,z \, .
\end{equation}

To consider the parity-doublet pattern of excited baryons, a pair of 5D spinors $N_1$ and $N_2$ corresponding to the spin-$\frac{1}{2}$ chiral baryon operators $\mathcal{O}_L$ and $\mathcal{O}_R$ in 4D shall be introduced as discussed in ref.\cite{Hong:2006ta}. The chiral baryon operators $\mathcal{O}_L$ and $\mathcal{O}_R$  transform as $(2,1)$ and $(1,2)$ under $SU(2)_L \times SU(2)_R$ respectively. The bulk action of the Dirac fields $N_j (j=1,2)$ is constructed as follows
\begin{equation}\label{Nj action}
S_{N_j} =\int d^5x \sqrt{g} \left[\frac{i}{2}\bar{N_j}e_A^M\Gamma^A\nabla_M N_j - \frac{i}{2}(\nabla_M^\dagger \bar{N_j})e_A^M\Gamma^A N_j - \hat{m}_N (z) \delta_j  \bar{N_j} N_j\right] \, ,
\end{equation}
where $\Gamma^A=(\gamma^{\mu},-i\gamma^5)$ are the 5D Dirac matrices which satisfy $\left\{\Gamma^A,\Gamma^B\right\}=2\eta^{AB}$, $e_A^M$ is the vielbein satisfying $g_{MN}=e_M^A e_N^B\eta_{AB}$, and $\nabla_M$ is the Lorentz and gauge covariant derivative
\begin{equation}\label{}
\nabla_M=\partial_M - \frac{i}{2}\omega_M^{AB}\Gamma_{AB} - i(A_L^a)_M t^a \, ,
\end{equation}
with $\omega_M^{AB}$ the spin connection given by $\omega_M^{AB}= \partial_z A(z)(\delta_M^A\delta_z^B-\delta_z^A\delta_M^B)$ and $\Gamma^{AB}=\frac{i}{4}\left[\Gamma^A,\Gamma^B\right]$.

Unlike the usual soft-wall AdS/QCD model for mesons in which there is an explicit dilaton term in the bulk action, such an exponential term of dilaton for the spin-$\frac{1}{2}$ fermion case can be removed from the action by making a rescaling definition for the fermionic baryon fields\cite{Gutsche:2011vb}. $\hat{m}_N(z)$ is the IR-modified 5D conformal mass of the bulk spinors
\begin{equation}
\hat{m}_N(z) = m_5+\tilde{m}_N(z)\, ,
\end{equation}
with $m_5$ the 5D conformal mass. The magnitude of $m_5$ is fixed by the general AdS/CFT relation with the scaling dimension $\Delta$ of the boundary operator\cite{Henningson:1998cd,Contino:2004vy}
\begin{equation}
m_5^2=\left(\Delta-\frac{d}{2}\right)^2 \, , \quad \Delta=\frac{9}{2} \, .
\end{equation}
Here we take $m_5=\frac{5}{2}$ and introduce $\delta_j=\pm1$ to yield the right chiral zero modes when matching $N_j$ with the 4D chiral operators $\mathcal{O}_L$ and $\mathcal{O}_R$ \cite{Hong:2006ta}
\begin{equation}
\delta_j=
\begin{cases}
1,  & j=1\\
-1, & j=2
\end{cases} \, .
\end{equation}
The IR-modified 5D mass term $\tilde{m}_N(z)$ is considered to have the following simple form
\begin{equation}
\tilde{m}_N(z)=\frac{\mu_g^2\,z^2}{1+\mu_g^2\,z^2}\, \lambda_N  \, ,
\end{equation}
which is required to be vanishing in the ultraviolet (UV) region $z\to 0$ and get a non-zero fixed value in the IR region $z\to \infty$ due to the QCD confining effect. Here $\mu_g$ is the energy scale characterizing the low energy QCD \cite{Cui:2013xva}. Note that such an IR behavior is different from a simple quadratic term appearing in \cite{deTeramond:2012rt,Gutsche:2011vb}.

Let us now construct the IR-modified Yukawa coupling term between the bulk spinors and bulk scalar field $X$. The 5D spinors $N_1$ and $N_2$ corresponding to the spin-$\frac{1}{2}$ chiral baryon operators $\mathcal{O}_L$ and $\mathcal{O}_R$ in 4D\cite{Hong:2006ta,Kim:2007xi}, it has the following form in 5D space-time
\begin{equation}\label{two flavors action coupling}
S_{Y} =\int d^5x \sqrt{g}\left[-y_N(z) (\bar{N}_1 X N_2+\bar{N}_2 X^\dagger N_1)\right] \, ,
\end{equation}
so that the resulting effective 4D Lagrangian possesses the chiral symmetry  $SU(N_f)_L \times SU(N_f)_R $ with $N_f$ the flavor number. Note that when the bulk vacuum expectation value (bVEV) of $X$ breaks the chiral symmetry, it remains keeping a vector-like symmetry, i.e., $SU(N_f)_L \times SU(N_f)_R \rightarrow SU(N_f)_V$. The IR-modified Yukawa coupling $y_N(z)$
is assumed to have the following properties at IR and UV regions
\begin{equation}
y_N(z)|_{z\to 0} \to 0 \, ; \qquad y_N(z) \langle X(z)\rangle|_{z\to \infty}   \to z^2 \, .
\end{equation}
It will be shown that the $S_{Y}$ term will lift the would-be massless ground state of baryons to a massive state and split the degenerate massive excitations into parity doublets, so that it can explain the experimental sign of the parity doubling pattern of excited baryons\cite{Jaffe:2006jy}. Note that the Yukawa coupling $y_N(z)$ is considered to be an IR-modified one which is different from the case in\cite{Hong:2006ta}, its form will be specified later as it is relevant to the bVEV of the bulk meson field.

The general action for the bulk baryon field is given by
\begin{equation}\label{two flavors action}
S_N=S_{N_1}+S_{N_2}+S_{Y}\, .
\end{equation}
We shall investigate the phenomena of resonance baryons based on the above action.

\subsection{Bulk Vacuum Expectation Value of  Scalar Meson Field}

The 5D action of the meson sector in the original soft-wall model \cite{Karch:2006pv} is given as

\begin{equation}\label{meson action}
S=\int d^{5}x\,\sqrt{g}e^{-\Phi(z)}\,{\rm {Tr}}\left[|DX|^{2}-m_{X}^{2} |X|^{2} - \frac{1}{4g_{5}^2}(F_{L}^2+F_{R}^2)\right]
\end{equation}
with $D^MX=\p^MX-i A_L^MX+i X A_R^M$, $A_{L,R}^M=A_{L,R}^{M~a}t^a$ and ${\rm{Tr}}[t^at^b]=\delta^{ab}/2$. The gauge coupling $g_5$ is fixed to be $g_5^2 = 12\pi^2/N_c$ with $N_c$ the color number \cite{Erlich:2005qh}. The complex bulk fields $X$ are in general decomposed into the  scalar and pseudoscalar mesons, the chiral gauge fields $A_L$ and $A_R$  are identified to the vector and axial-vector mesons.

The bVEV of the scalar meson field X has the form for two flavor case
\begin{equation}\label{VEV two flavor}
\chi(z)\equiv\VEV{X}=\frac{1}{2}v(z)\left(
                         \begin{array}{cc}
                           1 & 0 \\
                           0 & 1 \\
                         \end{array}
                       \right)
\end{equation}
which satisfies the equation of motion derived from the action (\ref{meson action})
\begin{equation}\label{salar field eqution1}
\chi''(z)+(3 A'(z)-\Phi'(z))\, \chi'(z)-m_X^2\, e^{2A(z)}\, \chi(z) = 0 \, .
\end{equation}
As expected from the AdS/CFT dictionary shown in \cite{Klebanov:1999tb,Witten:1998qj}, the bVEV $v(z)$  has the following behavior at the UV boundary $z\to 0$:
 \begin{equation} \label{vev}
 v(z\to0)=m_q \,\zeta\, z+\frac{\sigma\, z^3}{\zeta}\, ,
 \end{equation}
where $m_q$ and $\sigma$ are the current quark mass and quark condensate, respectively, and $\zeta=\sqrt{3}/(2\pi)$ is the normalization parameter\cite{Cherman:2008eh}.

In the original soft-wall model \cite{Karch:2006pv}, $v(z)$ can be calculated from the equation of motion Eq.(\ref{salar field eqution1}), but the solution will reach to a constant in the IR limit $z \rightarrow \infty$. Such an asymptotic behavior leads to the chiral symmetry restoration which is not supported in QCD\cite{Shifman:2007xn}. In the recent studies\cite{Gherghetta:2009ac,Sui:2009xe,Cui:2013xva}, the asymptotic behavior of $v(z)$ in the IR limit is assumed to have a reliable form which can realize the chiral symmetry breaking. Where the simplest forms for the dilaton field and the metric or the conformal mass have been modified in the IR region to make Eq. (\ref{salar field eqution1}) consistent.

As shown in ref.\cite{Cui:2013xva}, a consistent IR-improved soft-wall AdS/QCD model for mesons with the quartic interaction term can be written as
\begin{equation}
S=\int d^{5}x\,\sqrt{g}e^{-\Phi(z)}\,{\rm {Tr}}\left[|DX|^{2}-m_{X}^{2}(z) |X|^{2}-\lambda_X(z) |X|^{4}-\frac{1}{4g_{5}^2}(F_{L}^2+F_{R}^2)\right],\label{action}
\end{equation}
where the conformal mass $m_X(z)$ and the quartic coupling $\lambda_X(z)$ are considered to have the proper IR-improved forms. The bVEV $v(z)$ has the following IR behavior
\begin{equation} \label{vinfty}
 v(z\to \infty)= v_q z
 \end{equation}
with $v_q$ the constant parameter which characterizes the energy scale of dynamically generated spontaneous chiral symmetry breaking caused by the quark condensate \cite{DW}. A simple parameterized form of $v(z)$ is taken as\cite{Cui:2013xva}
\begin{equation}\label{bVEV}
    v(z)=\frac{A z+B z^3}{1+C z^2}.
\end{equation}
with
\begin{equation}\label{ABC}
     A=m_q\zeta,\quad B=\frac{\sigma}{\zeta}+m_q\zeta C,\quad C=\mu_c^2/\zeta,\quad v_q =B/C
\end{equation}
where the constant parameter $\mu_c$ characterizes the QCD confinement scale.

Then we specify the Yukawa coupling term $y_N(z)$ as
\begin{equation}\label{bVEV}
    y_N(z)= \lambda_A\, \mu_g z\, \left(1 - \lambda_B\, \mu_g^2 z^2 e^{-\mu_g^2 z^2} \right)\, ,
\end{equation}
which is required to realize a consistent mass spectrum of resonance baryons with experimental data.

\subsection{Equation of Motion and Solutions for Parity-doublet Baryons}

Let us decompose the bulk baryon fields $N_j(j=1,2)$ into the chiral form
\begin{equation} \label{N}
N_j=N_{jL}+N_{jR}
\end{equation}
where $i\Gamma^5 N_{jL}=N_{jL}$ and $i\Gamma^5 N_{jR}=-N_{jR}$. Then a KK decomposition for $N_{jL,R}$ is performed to yield the following form
\begin{equation} \label{fLR}
N_{jL,R}(x,z)=\sum_n\int\frac{d^4p}{(2\pi)^4}e^{-ipx}f_{jL,R}^{(n)}(z)\psi_{L,R}^{(n)}(p) \, ,
\end{equation}
where $\psi_{L,R}^{(n)}(p)$ is the 4D spinors satisfied with $\gamma^5 \psi_L^{(n)}(p)=\psi_L^{(n)}(p),\,\gamma^5 \psi_R^{(n)}(p)=-\psi_R^{(n)}(p)$ and $\slashed{p}\,\psi_{L,R}^{(n)}(p)=|p|\psi_{R,L}^{(n)}(p)$.

From the action given in Eq.(\ref{two flavors action}),  we can derive the equation of motion in terms of $f_{jL,R}\,(j=1,2)$ as (note that we shall neglect the superscript of $f_{jL,R}^{(n)}$ and $\psi_{L,R}^{(n)}$ below for convenience)
\begin{eqnarray}\label{nucleon equation}
&\begin{pmatrix} \p_z-e^{A(z)}\, \hat{m}_N(z)+\frac{d}{2}\,A^\prime(z)  &   -\,y_N\,e^{A(z)}\,\chi(z)\\   -\,y_N\,e^{A(z)}\,\chi(z)  &   \p_z+e^{A(z)}\, \hat{m}_N(z) +\frac{d}{2}\,A^\prime(z) \end{pmatrix}\,\begin{pmatrix} f_{1L}\\f_{2L}\\ \end{pmatrix}=-|p|\,\begin{pmatrix} f_{1R}\\f_{2R} \end{pmatrix}\nonumber \\
&\begin{pmatrix} \p_z+e^{A(z)}\, \hat{m}_N(z) +\frac{d}{2}\,A^\prime(z) &  y_N\,e^{A(z)}\,\chi(z)\\  y_N\,e^{A(z)}\,\chi(z)   &   \p_z-e^{A(z)}\, \hat{m}_N(z) +\frac{d}{2}\,A^\prime(z))\\ \end{pmatrix}\,\begin{pmatrix} f_{1R}\\f_{2R}\\ \end{pmatrix}=|p|\,\begin{pmatrix} f_{1L}\\f_{2L}\\ \end{pmatrix}  \, .
\end{eqnarray}

Here we need to clarify the even-odd parity of baryons, which is hidden in the above equation of motion. The parity transformations of 4D spinors $\psi_{L,R}$ are defined as (P is a unitary representation of parity transformation)
\begin{equation}\label{4D chiral parity}
P^{-1} \psi_{L,R}(x) P=\gamma^0 \psi_{R,L}(\bar{x}) \, .
\end{equation}
The parity transformation for the 5D spinors are defined as
\begin{equation}
\begin{aligned} \label{5D chiral parity}
P^{-1} N_{1L}(x,z) P = \eta_1 \gamma^0 N_{2R}(\bar{x},z),\qquad P^{-1} N_{1R}(x,z) P = \eta_2 \gamma^0 N_{2L}(\bar{x},z),\\
P^{-1} N_{2L}(x,z) P = \eta_2 \gamma^0 N_{1R}(\bar{x},z),\qquad P^{-1} N_{2R}(x,z) P = \eta_1 \gamma^0 N_{1L}(\bar{x},z).
\end{aligned}
\end{equation}

It is easy to prove that the action in Eq. (\ref{two flavors action}) is parity invariant if $\eta_1,\, \eta_2$ satisfy the relations \cite{He:2013gta,Gutsche:2012bp}
\begin{equation}\label{parity condition1}
\eta_1^* \eta_1=\eta_2^* \eta_2=1,\quad \eta_1^* \eta_2=\eta_2^* \eta_1=-1 \, .
\end{equation}
For convenience, we restrict $\eta_1$ and $\eta_2$ to be real. we will show that $\eta_1=1,\eta_2=-1$ correspond to the even parity case, while $\eta_1=-1,\eta_2=1$ correspond to the odd parity case. Using Eq.(\ref{4D chiral parity}) and Eq.(\ref{5D chiral parity}), we can easily find the relations between the bulk profiles $f_{jL,R} (j=1,2)$ corresponding to the even and odd parity, respectively
\begin{gather}\label{parity condition2}
\begin{aligned}
f_{1L}(z)&=f_{2R}(z),\quad f_{1R}(z)=-f_{2L}(z)\qquad \text{even},\\
f_{1L}(z)&=-f_{2R}(z),\quad f_{1R}(z)=f_{2L}(z)\qquad \text{odd}.
\end{aligned}
\end{gather}

Now we introduce $f_j^+(z)$ and $f_j^-(z)$ as

\begin{equation}\label{f+-}
\begin{aligned}
f_1^+(z) &= f_{1L}(z)+f_{2R}(z),& \qquad f_2^+(z) &= f_{1R}(z)-f_{2L}(z),\\
f_1^-(z) &= f_{1L}(z)-f_{2R}(z),& \qquad f_2^-(z) &= f_{1R}(z)+f_{2L}(z).
\end{aligned}
\end{equation}

Then Eq.(\ref{nucleon equation}) can be decoupled as
\begin{eqnarray}\label{f+ equation}
&\begin{pmatrix} -\p_z+e^{A(z)}\, \hat{m}_N(z) -\frac{d}{2}\,A^\prime(z)  &   -\,y_N\,e^{A(z)}\,\chi(z)\\   \,y_N\,e^{A(z)}\,\chi(z)  &   \p_z+e^{A(z)}\, \hat{m}_N(z) +\frac{d}{2}\,A^\prime(z) \end{pmatrix}\,\begin{pmatrix} f_1^+\\f_2^+\\ \end{pmatrix}=|p|\,\begin{pmatrix} f_2^+\\f_1^+ \end{pmatrix} \, ,
\end{eqnarray}
\begin{eqnarray}\label{f- equation}
&\begin{pmatrix} -\p_z+e^{A(z)}\, \hat{m}_N(z) -\frac{d}{2}\,A^\prime(z)  &   \,y_N\,e^{A(z)}\,\chi(z)\\   -\,y_N\,e^{A(z)}\,\chi(z)  &   \p_z+e^{A(z)}\, \hat{m}_N(z) +\frac{d}{2}\,A^\prime(z) \end{pmatrix}\,\begin{pmatrix} f_1^-\\f_2^-\\ \end{pmatrix}=|p|\,\begin{pmatrix} f_2^-\\f_1^- \end{pmatrix} \, .
\end{eqnarray}
From Eq.(\ref{parity condition2})-(\ref{f+-}), we can see that for the even parity case only  $f_j^+ (j=1,2)$ survive and for the odd parity case only  $f_j^- (j=1,2)$ survive. $f_j^+$ and $f_j^-$ are just the holographic analogues of the even and odd parity baryon wave functions, respectively.

\subsection{Numerical Results}

In the numerical calculations, we consider for simplicity only two flavor case with lightest quarks $u$ and $d$ or nucleons $p$ and $n$ and ignore the isospin symmetry breaking. The model involves seven parameters $m_q$, $\sigma$, $\mu_g$, $\mu_c$, $\lambda_N$, $\lambda_A$ and $\lambda_B$. While four of the parameters appear in the AdS/QCD model for mesons, i.e., quark mass $m_q$, quark condensate $\sigma$, energy scales $\mu_c$ and $\mu_g$. Here we shall take their values obtained from the IR-improved soft-wall AdS/QCD model for mesons as it was shown to provide a consistent prediction for the meson mass spectra of resonance scalars, pseudoscalars, vectors and axial-vectors\cite{Cui:2013xva}. Thus there are only three parameters $\lambda_N$, $\lambda_A$ and $\lambda_B$ in the IR-improved soft-wall AdS/QCD model for nucleons, which are determined by taking a global fitting for nucleon mass spectrum. The input values are shown in the Table \ref{Table:parameter}

\begin{table}[!h]
\begin{center}
\begin{tabular}{cccccccc}
\hline\hline
$m_q(\MeV)$    &   $\sigma^{1/3}(\MeV)$    &    $\mu_g(\MeV)$     &    $\mu_c(\MeV)$     &   $\lambda_A$    &    $\lambda_B$    &    $\lambda_N$   \\
\hline
3.52           &           290             &        473           &        375         &    3.93   &   16.58    &      2.55       \\
\hline\hline
\end{tabular}
\caption{The values of input parameters. $m_q$, $\sigma$, $\mu_g$, $\mu_c$ are taken from \cite{Cui:2013xva}. }
\label{Table:parameter}
\end{center}
\end{table}

The eigenvalues of the nucleons masses can be calculated by solving the equation of motion Eq.(\ref{f+ equation}) and Eq.(\ref{f- equation}) numerically with the boundary conditions $f_1^+(z\rightarrow0)=f_2^+(z\rightarrow0)=0$ and $f_1^-(z\rightarrow0)=f_2^-(z\rightarrow0)=0$. The results are presented in the Table \ref{Table:nucleon even mass} and \ref{Table:nucleon odd mass}. One can see a good agreement between the theoretical calculations and experimental measurements, which is much better than the hard-wall results in \cite{Hong:2006ta}.

\begin{table}[!h]
\begin{center}
   \begin{tabular}{|c|c|c|c|c|c|c|c|}
     \hline
     $N (even)$ & 0 & 1 & 2 & 3 & 4 & 5 \\
     \hline\hline
     Exp.(MeV) & 939 **** & 1440**** & 1710 *** & 1880  ** & 2100   *  & 2300  **  \\
     \hline
     Theory(MeV) & 939 & 1435 & 1698 & 1915 & 2105 & 2276 \\
     \hline
   \end{tabular}
\end{center}
\caption{Mass spectrum of even parity nucleons. The superscript $\ast$ represents the existence confidence level. The experimental data are from \cite{Agashe:2014kda}. }
\label{Table:nucleon even mass}
\end{table}

\begin{table}[!h]
\begin{center}
\begin{tabular}{|c|c|c|c|c|c|c|c|}
     \hline
     $N (odd)$ & 0 & 1 & 2 & 3 & 4 \\
     \hline\hline
     Exp.(MeV) & 1535**** & 1650**** & 1895  ** & ----- & -----  \\
     \hline
     Theory(MeV) & 1473 & 1717 & 1927 & 2113 & 2281 \\
     \hline
\end{tabular}
\end{center}
\caption{Mass spectrum of odd parity nucleons. The superscript $\ast$ represents the existence confidence level. The experimental data are from \cite{Agashe:2014kda}. }
\label{Table:nucleon odd mass}
\end{table}

\begin{figure}[!h]
\begin{center}
\includegraphics[width=64mm,clip]{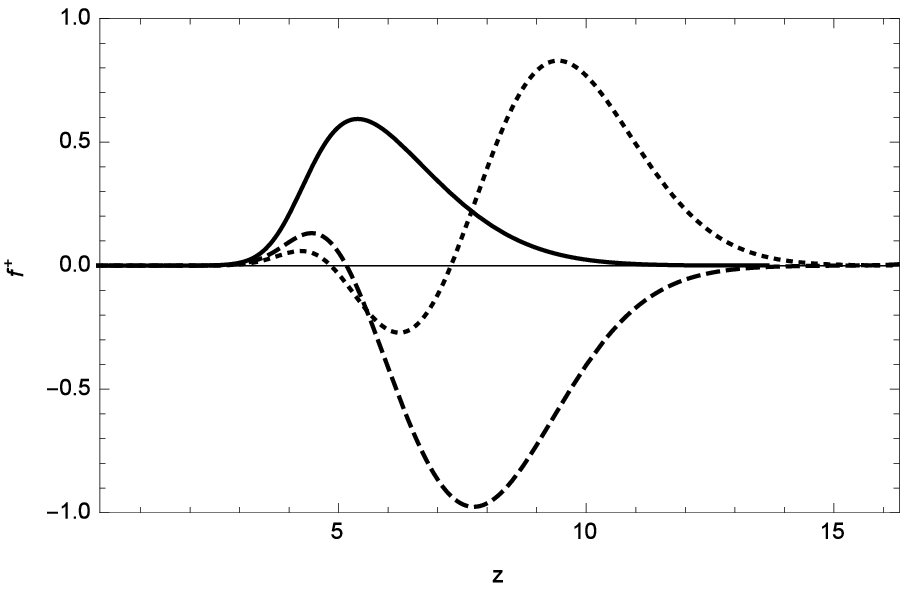}
\includegraphics[width=64mm,clip]{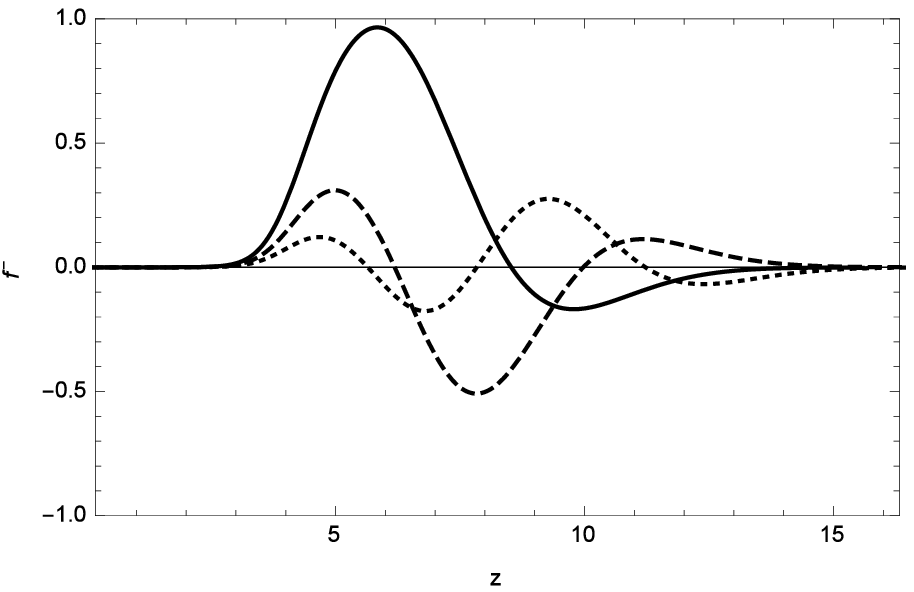}
\end{center}
\caption{baryon wave functions for the first three even parity states(left) and odd parity states (right). The solid line represents ground states, the dashed line first excited states, and the dotted line second excited states.}\label{baryon wave functions}
\end{figure}

We plot the baryon wave functions in Fig.\ref{baryon wave functions}.

\section{(Axial) vector meson-nucleon couplings}

It is useful to apply the IR-improved AdS/QCD model for baryons described above to calculate the $\rho(a_1)$ meson-nucleon couplings. Note that many efforts have been made in the coupling constant calculation. The vector meson-nucleon coupling constants have been evaluated in the top-down approach \cite{Hong:2005np,Hong:2007ay,Hashimoto:2008zw} and in the framework of hard-wall model \cite{Ahn:2009px,Maru:2009ux}. There are also empirical data from experiments or other effective models \cite{Hohler:1974ht,Machleidt:2000ge,Gross:2007be,Stoks:1996yj,Aliev:2009ei,Riska:2000gd chiral quark model}.

As was shown in \cite{Hong:2007tf,Maru:2009ux}, the vector-nucleon couplings may receive contributions in two ways. The first one comes from the gauge interactions which incorporate in the covariant derivative of the kinetic term of the action Eq.(\ref{Nj action}),
\begin{eqnarray}\label{gauge term}
S^{gauge}
&=& \int d^5x \sqrt{g} \left[\frac{i}{2}\bar{N_1}e_A^M\Gamma^A (-i A_M^L) N_1 - \frac{i}{2}((-i A_M^L) N_1)^\dagger \Gamma^0 e_A^M\Gamma^A N_1 \right. \nonumber \\
& & \left. +\frac{i}{2}\bar{N_2}e_A^M\Gamma^A (-i A_M^R) N_2 - \frac{i}{2}((-i A_M^R) N_2)^\dagger \Gamma^0 e_A^M\Gamma^A N_2\right] \nonumber  \\
&\supset& \int d^4x \int_{0}^{\infty} \frac{dz}{z^4} \left[\bar{N_1} \gamma^{\mu} V_{\mu} N_1 + \bar{N_2} \gamma^{\mu} V_{\mu} N_2 + \bar{N_1} \gamma^{\mu} A_{\mu} N_1 - \bar{N_2} \gamma^{\mu} A_{\mu} N_2\right]  \nonumber \\
&=& \int d^4x \int_{0}^{\infty} \frac{dz}{z^4} \left[V_0(z)(|f_{1L}^{(0)}(z)|^2+|f_{2L}^{(0)}(z)|^2)\bar{\psi} \gamma^{\mu} V_{\mu}(x) \psi \right. \nonumber \\
& & \left. - A_0(z)(|f_{1L}^{(0)}(z)|^2-|f_{2L}^{(0)}(z)|^2)\bar{\psi} \gamma^5 \gamma^{\mu} A_{\mu}(x) \psi \right] \, .
\end{eqnarray}

The second one originates from the Pauli term which was introduced in \cite{Hong:2007tf} for the evaluation of the anomalous magnetic dipole moment $\mu_{ano}$ and the CP-violating electric dipole moment $d_e$.
\begin{eqnarray}\label{Pauli term}
S^{Pauli}
&=& i\,\kappa\int d^5x \sqrt{g} e_A^M e_B^N \left[\bar{N_1}\Gamma^{AB}(F_L)_{MN} N_1 - \bar{N_2}\Gamma^{AB}(F_R)_{MN} N_2\right]  \nonumber \\
&\supset& i\,\kappa\int d^5x \sqrt{g} z^2 \left[\bar{N_{1L}}\Gamma^{\mu z}(F_V)_{\mu z} N_{1L} + \bar{N_{1R}}\Gamma^{\mu z}(F_V)_{\mu z} N_{1R} - (1 \leftrightarrow 2)\right] \nonumber  \\
& & i\,\kappa\int d^5x \sqrt{g} z^2 \left[\bar{N_{1L}}\Gamma^{\mu z}(F_A)_{\mu z} N_{1L} + \bar{N_{1R}}\Gamma^{\mu z}(F_A)_{\mu z} N_{1R} + (1 \leftrightarrow 2)\right] \nonumber \\
&=& -2\,\kappa \int d^4x \int_{0}^{\infty} \frac{dz}{z^3} \left[V_0^{\prime}(z)(|f_{1L}^{(0)}(z)|^2-|f_{2L}^{(0)}(z)|^2)\bar{\psi} \gamma^{\mu} V_{\mu}(x)\psi \right. \nonumber \\
& & \left. - A_0^{\prime}(z)(|f_{1L}^{(0)}(z)|^2+|f_{2L}^{(0)}(z)|^2)\bar{\psi} \gamma^5 \gamma^{\mu} A_{\mu}(x) \psi\right] \, .
\end{eqnarray}

We can easily read the $\rho(a_1)$ meson-nucleon coupling constants from the above two formulas
\begin{eqnarray}\label{gvnn}
g_{\rho NN}
&=& g^{(0)}_{\rho NN} + g^{(1)}_{\rho NN} \nonumber  \\
&=& \int_{0}^{\infty} \frac{dz}{z^4}\,V_0(z)\,(|f_{1L}^{(0)}(z)|^2+|f_{2L}^{(0)}(z)|^2) \nonumber \\
& & -2 \kappa \int_{0}^{\infty} \frac{dz}{z^3}\,V^\prime_0(z)\,(|f_{1L}^{(0)}(z)|^2-|f_{2L}^{(0)}(z)|^2) \, ,
\end{eqnarray}
and
\begin{eqnarray}\label{gann}
g_{a_1 NN}
&=& g^{(0)}_{a_1 NN} + g^{(1)}_{a_1 NN} \nonumber  \\
&=& \int_{0}^{\infty} \frac{dz}{z^4}\,A_0(z)\,(|f_{1L}^{(0)}(z)|^2-|f_{2L}^{(0)}(z)|^2) \nonumber  \\
& & -2 \kappa \int_{0}^{\infty} \frac{dz}{z^3}\,A^\prime_0(z)\,(|f_{1L}^{(0)}(z)|^2+|f_{2L}^{(0)}(z)|^2) \, .
\end{eqnarray}

Note that there are some differences of the above formulas from that in \cite{Maru:2009ux} though we follow the same procedure with the same action terms. The normalized wave function $(A_0) V_0$ of $(a_1) \rho$ meson has been determined in the IR-improved AdS/QCD model for mesons\cite{{Cui:2013xva}},  and $\kappa$ is fixed by the anomalous magnetic dipole moments of nucleons\cite{Hong:2007tf}
\begin{equation}\label{anomalous magnetic moment}
\mu_{ano} = -e\, 2\, \kappa \int_{0}^{\infty} \frac{dz}{z^3} f_{1L}^{(0)}(z) f_{2L}^{(0)}(z) \simeq \frac{1.8 e}{2 m_N}\, .
\end{equation}
With the nucleon mass $m_N \simeq 0.939 GeV$, we get $\kappa \simeq 0.19$.

Taking the parameters obtained in the IR-improved AdS/QCD model for mesons\cite{{Cui:2013xva}} and the IR-improved AdS/QCD model for baryons in this paper, we are able to calculate the $\rho(a_1)$ meson-nucleon coupling constant. The result is shown in the Table \ref{Table:vnnanncoupling} in comparison with other models or empirical values. Where \cite{Hohler:1974ht,Machleidt:2000ge,Gross:2007be,Stoks:1996yj} are empirical estimations from different experimental data, and \cite{Aliev:2009ei} is a QCD sum rule calculation. The result of \cite{Riska:2000gd chiral quark model} was obtained in the framework of chiral quark model. \cite{Maru:2009ux,Ahn:2009px} are hard-wall models with different action terms. \cite{Huseynova:2014pca} is about the calculation of the $\rho$ meson-nucleon coupling in a soft-wall AdS/QCD model which is different from our present one. In our calculations, the $\rho$ and $a_1$ meson-nucleon coupling constants have included two contributions from both the gauge and Pauli action terms. The numerical results of these terms are: $ g_{\rho NN}^{(0)}\simeq2.71$, $ g_{\rho NN}^{(1)}\simeq-0.23$, $ g_{a_1 NN}^{(0)}\simeq0.29$, $ g_{a_1 NN}^{(1)}\simeq-0.15$. It is seen that the gauge interactions give the main contribution of the couplings, while the Pauli term contributes to a small negative part.

\begin{table}[!h]
\begin{center}
\begin{tabular}{ccc}
\hline
Model/Experiment               &              $ g_{\rho NN}$              &             $g_{a_1 NN}$     \\[1ex]
\hline
Present model                &                   2.48                   &                 0.14         \\
experiment\cite{Hohler:1974ht,Machleidt:2000ge,Gross:2007be}& 4.2$\sim$6.5  &                --        \\
experiment\cite{Stoks:1996yj}    &                 2.52 $\pm$ 0.06          &                --       \\
sum rule\cite{Aliev:2009ei}    &               -2.5 $\pm$ 1.1             &               --       \\
chiral quark\cite{Riska:2000gd chiral quark model}&    2.8                    &                  --       \\
\hline
hard-wall\cite{Maru:2009ux}     &                  0.2 $\sim$ 0.5          &          1.5 $\sim$ 4.5         \\
hard-wall\cite{Ahn:2009px}      &                 -3.42 (-8.6)             &               --       \\
soft-twall\cite{Huseynova:2014pca}&                5.33 (6.78)              &        --      \\
\hline
\end{tabular}
\caption{The values of $\rho(a_1)$ meson-nucleon coupling constants.}
\label{Table:vnnanncoupling}
\end{center}
\end{table}

\newpage

\section{Conclusions and Remarks}
\label{Chap:Sum}

We have built an IR-improved AdS/QCD model for baryons. Two bulk spin-$\frac{1}{2}$ fermion fields and a bulk scalar field have been introduced to incorporate the effects of chiral symmetry breaking which is crucial for the low mass baryons. The Yukawa coupling term in the action has induced the chiral symmetry breaking and split the baryons into a parity-doublet pattern of resonance states. The IR-modified 5D conformal mass $\tilde{m}_N(z)$ plays the role of an effective confining potential for obtaining the reliable mass spectrum of baryons. By adopting the parameterization for the bulk scalar field resulted from the IR-improved soft-wall AdS/QCD model for mesons\cite{Cui:2013xva}, we have arrived at a consistent mass spectrum of baryons, which agrees well with the experimental data. It has been shown that the combined behavior of the bVEV of the bulk scalar field and the IR-modified Yukawa coupling $y_N(z)$ is critical for yielding the consistent mass spectra of the highly excited baryon states.

The vector-nucleon coupling constants $g_{\rho NN}$ and $g_{a_1 NN}$ have been calculated within the IR-improved AdS/QCD model for baryons. We have considered both the gauge interaction term contained in the covariant derivative of action Eq.(\ref{Nj action}) and the terms related to the anomalous magnetic dipole moment of nucleons \cite{Hong:2007tf,Maru:2009ux}.  The numerical result of $g_{\rho NN}\simeq2.48$ is consistent with the experimental data and many other models from QCD. The coupling $g_{a_1 NN}$ has no experimental data, our resulting value is much smaller than the one obtained in the hard-wall model\cite{Maru:2009ux} though the same action terms are taken. Note that the formula of $g_{\rho NN}$ and $g_{a_1 NN}$ are different from that in \cite{Maru:2009ux}.

In this paper we have carried out the calculations for the mass spectrum of baryons for two flavour case and discussed the vector-nucleon coupling constants. In general, many other properties relevant to baryons can be studied within the framework of the IR-improved AdS/QCD model for baryons, such as the nucleon electromagnetic and gravitational form factors, the pion-nucleon coupling. In particular, the extension to three flavours is an interesting case to be investigated by considering the explicit chiral symmetry breaking due to quark masses.

\section*{Acknowledgements}

This work was supported in part by the National Science Foundation of China (NSFC) under Grant \#No. 11475237, No.~11121064, No.~10821504 and also by the CAS Center for Excellence in Particle Physics (CCEPP).


\end{document}